\documentclass[10pt,twocolumn,oneside]{IEEEtran}
\IEEEoverridecommandlockouts

\usepackage{graphicx}
\usepackage{hyperref}
\usepackage{subfigure}
\usepackage{color}
\usepackage{multicol}
\usepackage{microtype}
\usepackage{amssymb}
\usepackage{multirow}
\usepackage{amsmath}
\usepackage{bm}
\usepackage{cite}
\usepackage{gensymb}
\usepackage{amsfonts}
\usepackage{mathrsfs}
\usepackage{amsmath}
\usepackage{algorithm}
\usepackage{algorithmic}
\usepackage{amsthm}

\usepackage{tabularx}
\usepackage{balance}
\usepackage{mathrsfs}
\usepackage{array}
\usepackage{multirow}
\usepackage{caption}
\newcommand{\PreserveBackslash}[1]{\let\temp=\\#1\let\\=\temp}
\newcolumntype{C}[1]{>{\PreserveBackslash\centering}p{#1}}
\newcolumntype{L}[1]{>{\PreserveBackslash\raggedright}p{#1}}
\newcolumntype{R}[1]{>{\PreserveBackslash\raggedleft}p{#1}}
\usepackage[flushleft]{threeparttable}
\begin{document}

	\title{Near-Field Communications for 6G: Fundamentals, Challenges, 
	Potentials, and Future Directions}
	\author{ Mingyao Cui, Zidong Wu, Yu Lu, Xiuhong Wei, and Linglong Dai, 
	\emph{Fellow, IEEE}
		
	\thanks{
		All authors are with the Beijing National Research Center for 
		Information Science and Technology (BNRist) as well as the Department 
		of Electronic Engineering, Tsinghua University, Beijing 100084, China 
		(e-mails: \{cmy20, wuzd19, y-lu19, weixh19\}@mails.tsinghua.edu.cn, 
		daill@tsinghua.edu.cn).
	
		© 2022 IEEE. Personal use of this material is permitted. Permission 
		from IEEE must be obtained for all other uses, in any current or future 
		media, including reprinting/republishing this material for advertising 
		or promotional purposes, creating new collective works, for resale or 
		redistribution to servers or lists, or reuse of any copyrighted 
		component of this work in other works.
}
}
	\maketitle
	\IEEEpeerreviewmaketitle
	\begin{abstract}
Extremely large-scale antenna array (ELAA) is a common feature of several key
candidate technologies for the sixth generation (6G) mobile networks, such as 
ultra-massive 
multiple-input-multiple-output (UM-MIMO), cell-free massive MIMO, 
reconfigurable intelligent surface (RIS), and terahertz communications.
{\color{black} Since the number of antennas is very large for ELAA, the 
electromagnetic radiation field needs to be modeled by near-field spherical 
waves, which differs from the conventional planar-wave-based radiation model of 
5G massive MIMO. As a result, near-field communications will become essential 
in 6G wireless networks.}
In this article, we systematically investigate the emerging near-field 
communication techniques. 
Firstly, we present the fundamentals of near-field 
communications and the metric to determine the near-field ranges 
in typical communication scenarios. 
Then, we investigate recent studies specific to near-field communications by 
classifying them into 
two categories, i.e.,  techniques addressing the challenges and those 
exploiting the potentials in near-field regions.
Their principles, recent progress, pros and cons are discussed. More 
importantly, several open 
problems and future research directions for near-field communications are 
pointed 
out. We believe that this article would inspire more innovations for this 
important research topic of near-field communications for 6G.
	\end{abstract}
		\begin{IEEEkeywords}
		6G, ELAA, near-field communications, spherical wavefront.\vspace{-3mm}
	\end{IEEEkeywords}

	\section{Introduction} \label{sec:1}

{\color{black} The sixth generation (6G) mobile networks} are promising to 
empower emerging applications, such as 
holographic video, digital replica, etc. 
For fulfilling these visions, tremendous research efforts have been endeavored 
to develop new wireless technologies to meet the key performance indicators 
(KPIs) of 6G, which are much superior to those of 5G \cite{6GVision_Saad2020}. 
For instance, thanks to the enormous spatial multiplexing and beamforming gain, 
ultra-massive multiple-input-multiple-output (UM-MIMO) and cell-free massive 
MIMO (CF-MIMO) are expected to accomplish a 10-fold increase in the spectral 
efficiency for 6G\cite{6GVision_Saad2020}. Besides, by dynamically manipulating 
the wireless environment through thousands of antennas, reconfigurable 
intelligent surface (RIS) brings new possibilities for capacity and coverage 
enhancement \cite{RFocus_MIT2020}.
{\color{black} Moreover, millimeter-wave (mmWave) and terahertz (THz) UM-MIMO can 
offer abundant spectral resources for supporting 100$\times$ peak data rate 
improvement (e.g., Tbps) in 6G mobile communications 
\cite{THzMIMO_Ian2016}.} Despite being suitable for different 
application scenarios with 
various KPIs, all the above technologies, including UM-MIMO, CF-MIMO, 
RIS, and THz communications,  
share a common feature: They all usually require a very large number of 
antennas to attain their expected performance, i.e.,  extremely large-scale 
antenna 
arrays (ELAA) are essential to these different candidate 
technologies for 6G.

Compared with massive MIMO, the key technology in 5G networks, ELAA for 6G not 
only 
means a sharp increase in the 
number of antennas but also results in a fundamental change of the 
electromagnetic (EM) characteristics.
The EM radiation field can generally be divided into far-field and radiation 
near-field 
regions. 
{\color{black} The boundary between these two regions is determined by the 
Rayleigh distance, also called the Fraunhofer distance 
\cite{Fresnel_Selvan2017}.}
Rayleigh distance is proportional to the product of the square of array aperture and carrier frequency~\cite{Fresnel_Selvan2017}.	
Outside the Rayleigh distance, it is the far-field region, where the EM field 
can be approximately modeled by \emph{planar waves}.
Within the Rayleigh distance, the near-field propagation becomes dominant, 
where the EM field has to be accurately modeled by 
\emph{spherical waves}. 

Since the number of antennas is not very large in 5G massive MIMO systems, the 
Rayleigh 
distance of up to several meters is negligible. 
Thus, existing 5G communications are mainly developed from far-field 
communication theories and techniques. 
However, with the significant increase of the antenna number and carrier 
frequency in future 6G systems, 
the near-field region of ELAA will expand by orders of magnitude. 
{\color{black}For instance, a 3200-element ELAA at 2.4 GHz was developed in \cite{RFocus_MIT2020}. With an array size of $2\:\text{m}\times 3\:\text{m}$, its Rayleigh distance is about 200 meters, which 
is larger than the radius of a typical 5G cell.} 
{\color{black}Accordingly, near-field communications will become essential 
components in future 6G mobile networks where the spherical propagation model 
needs to be 
considered, which is obviously different from the existing far-field 5G 
systems.}
Unfortunately, the near-field propagation 
introduces several new challenges in ELAA systems, which should 
be identified and addressed to empower 
6G communications.

In this article, we systematically investigate the recent near-field 
communication 
techniques for 6G. The key features of this article can be summarized as 
follows:
\begin{itemize}
	\item To begin with, the fundamental differences between  
	far-field and near-field communications are explained. 
	Comparatively speaking, the planar wavefront in the far-field can steer 
	the signal energy towards a specific physical \emph{angle}. On the 
	contrary,
	the near-field spherical wavefront achieves energy focusing on 
	\emph{both angle and distance} domain.
	Moreover, the Rayleigh distance that quantifies the boundary between far-field and near-field regions is introduced, and its derivation is explained in detail. 
	Based on this derivation, we further extend the classical Rayleigh distance, for MIMO channels with a direct base station (BS)-user equipment (UE) link, to the one for RIS-aided communications, where a cascaded channel is utilized for presenting the BS-RIS-UE link.
	\item Additionally, we investigate the emerging near-field communication 
	techniques 
	by classifying	them into two types, i.e., techniques addressing 
	the challenges and those exploiting the potentials in near-field 
	regions. On the one hand, as most techniques specific to far-field often suffer from a severe performance loss in the near-field area, the first type of techniques aims to compensate for this loss, such as near-field channel estimation and beamforming. 
	On the other hand, the second kind of study has 
	revealed that the nature of near-field spherical wavefront can also be 
	exploited to provide new possibilities for capacity 
	enhancement and accessibility improvement.  The principles, 
	recent progress, pros and cons of these two 
	categories of research are discussed in detail. 
	\item Finally, several open problems and future research directions for 
	near-field communications are pointed out. For example, the improvement of Rayleigh distance considering various communication metrics need to be analyzed, artificial intelligence (AI) is expected 
	to enable high-performance near-field transmissions with low complexity, 
	and hybrid far- and near-field communications also require in-depth study.
	
\end{itemize}

	\section{Fundamentals of Near-Field Communications} \label{sec:2}
In this section, we first present the differences between 
far-field and near-field communications. Then, we will identify the 
principle to determine the boundary between the far-field and near-field 
regions in several typical application scenarios.
	\subsection{Far-Field Communications vs. Near-Field Communications} 
	\label{sec:2-1}

\begin{figure}
		\centering
		%		\vspace*{-0.5em}
		\includegraphics[width=3.5in]{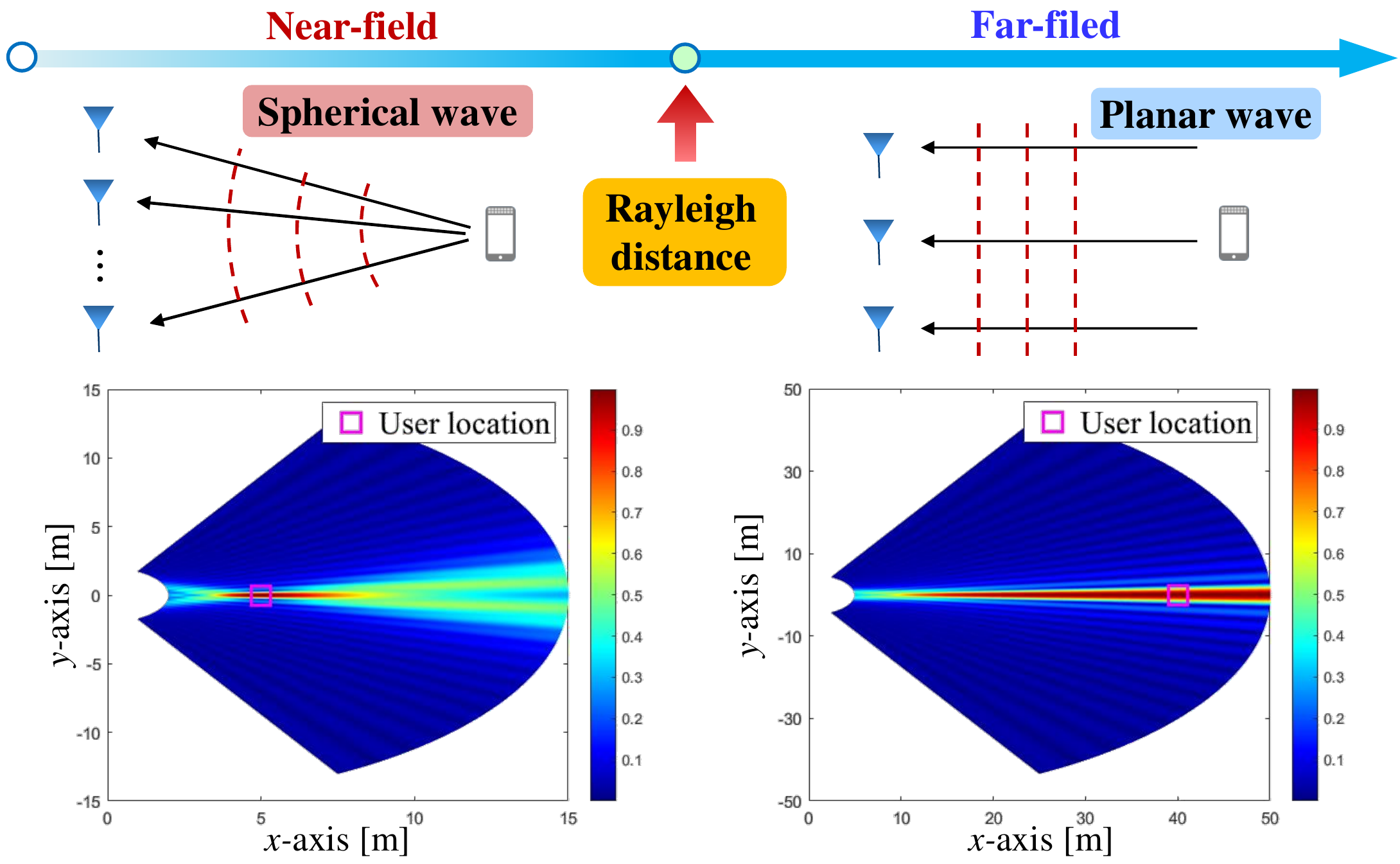}
		\vspace*{-1em}
		\captionsetup{font = {small}}
		\caption{ Far-field planar wavefront vs. near-field 
		spherical wavefront. {The plots at the bottom} illustrate the normalized 
		received signal energy in the physical space achieved by near-field 
		beamfocusing (bottom left) and far-field beamsteering (bottom right).
		}
		\label{fig1}
		\vspace*{-1em}
	\end{figure}

	The critical characteristics of far-field and near-field communications are 
	shown in Fig. \ref{fig1}. 	We consider an uplink communication scenario, 
	while the discussions in this article are also valid for downlink 
	scenarios. The BS is equipped with an ELAA.  {\color{black} A widely adopted 
	metric to determine the boundary between far-field and near-field regions 
	is the Rayleigh distance, also called the Fraunhofer distance 
	\cite{Fresnel_Selvan2017}.}
	When the communication distance between the BS and UE (BS-UE distance)  is 
	larger than the Rayleigh distance, the UE is located in the far-field 
	region of the BS. Then, EM waves impinging on the BS array can be 
	approximately modeled as planar waves.
	By contrast, when the BS-UE distance is shorter than the Rayleigh distance, 
	the UE is located in the near-field region of the BS. In this region, 
	EM waves impinging on the BS array must be accurately modeled as spherical 
	waves 
	\cite{Near_Cui2022}.

{\color{black}More precisely, the planar wave is a long-distance approximation of 
the spherical wave. In far-field regions, the phase of EM waves can be 
elegantly 
approximated by a {\bf{\itshape{linear}}} function of the antenna index through 
Taylor expansion. This concise linear phase forms a 
planar wavefront only related to an incident angle. Accordingly, by the 
utilization of planar wavefronts, far-field beamforming can steer the beam 
energy towards a  specific angle over different distances, which 
is also termed }{ \color{black}
as beamsteering, as shown in the bottom right figure of Fig. \ref{fig1}. 
Unfortunately, this concise linear phase fails to thoroughly  reveal the 
information of spherical waves.
In near-field regions, the phase of spherical waves should be accurately 
derived based on the physical geometry, which is a {\bf{\itshape{non-linear}}} 
function of the antenna index. The information of the incident angle and 
distance in each path between BS and UE is embedded in this non-linear phase. 
Exploiting the extra distance information of  spherical wavefronts,  near-field 
beamforming is able to focus the beam energy on a specific location, where 
energy focusing on both the angle and distance domain is achievable, as shown 
in the bottom 
left figure of Fig. \ref{fig1}. Owing to this property, beamforming in the 
near-field is also called  beamfocusing.}

The differences between far-field planar wavefronts and near-field spherical 
wavefronts bring several challenges and potentials to wireless communications, 
which will be detailed in the following sections.

	\subsection{Rayleigh Distance} \label{sec:2-2}
		\begin{figure*}
			% \begin{center}
		\centering
				\vspace*{-0.5em}
		\includegraphics[width=6.7in]{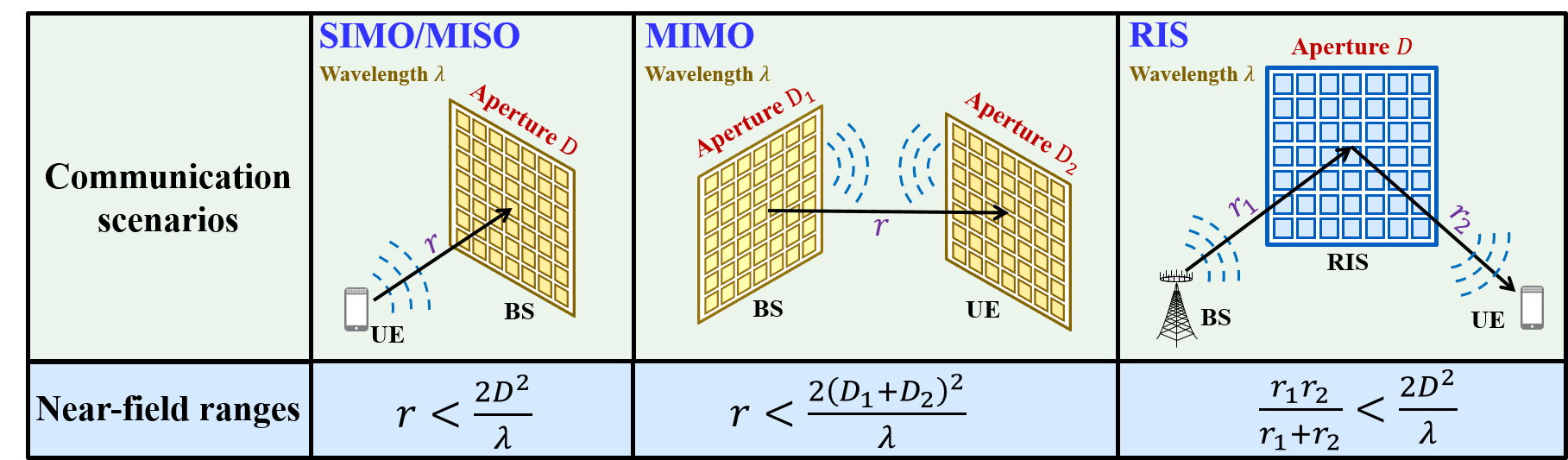}
		% \vspace*{-1em}
		\captionsetup{font = {small}}
		\caption{Near-field ranges for typical scenarios. 
		}
		\label{fig2}
		\vspace*{-1em}
			% \end{center}
	\end{figure*}

	The most crucial premise for near-field communications is quantifying
	the boundary between the far-field and near-field regions, i.e., the 
	Rayleigh distance.
 Generally, the classical Rayleigh distance is proportional to the square of 
 the array aperture and the inverse of the wavelength. 
Its derivation can be summarized as follows 
\cite{Fresnel_Selvan2017}. 
	The true phase of the EM wave impinging on a BS antenna has to be calculated based on the accurate spherical wave model.
In far-field scenarios,  this 
phase is usually approximated by its first-order Taylor expansion based on the 
planar wavefront model.
This approximation results in a phase discrepancy, which increases when the 
distance decreases.	 
When the largest phase discrepancy among all BS and UE antennas reaches 
${\pi}/{8}$, the distance between the BS array center and
the UE array center is defined as the Rayleigh distance. 
Accordingly, if the communication distance is shorter than the Rayleigh 
distance, the largest phase discrepancy will be larger than ${\pi}/{8}$. In this 
case, the far-field approximation becomes inaccurate, and thus  
the near-field propagation needs to be utilized.

	Based on this definition, the near-field ranges for 
	SIMO, MISO, and MIMO communication systems can be obtained. As illustrated 
	in Fig.	\ref{fig2}, the near-field range of SIMO/MISO scenarios is 
	precisely determined by the classical Rayleigh distance, which is 
	proportional to \textbf{\emph{the square of BS array aperture}}. 
	For the MIMO scenario, since ELAAs are employed at two sides of the BS-UE 
	link, both the BS array aperture and the UE array aperture contribute 
	to the Rayleigh distance, i.e., the near-field range is proportional to 
	\textbf{\emph{the square of the sum of BS array aperture and UE array 
	aperture}}.
	
Interestingly enough, we further extend the conventional Rayleigh 
	distance derived in SIMO/MISO/MIMO systems to that in RIS-aided 
	communication systems, as shown in Fig. \ref{fig2}.
Unlike SIMO/MISO/MIMO channels with a direct BS-UE link, the 
cascaded 
BS-RIS-UE channel in RIS systems comprises the BS-RIS and 
RIS-UE links. Therefore, when calculating 
phase discrepancy,  the BS-RIS distance
and the RIS-UE distance need to be added. 
Then, capturing the largest phase discrepancy of $\pi/8$, 
the near-field range in RIS systems is determined by \textbf{\emph{the harmonic 
mean of the BS-RIS distance
and the RIS-UE distance}}, as shown in Fig. \ref{fig2}. 
It can be further implied from Fig. \ref{fig2} that, as long as any of these two distances is shorter than the Rayleigh distance,
RIS-aided communication is operating in the near-field area. 
Therefore, near-field propagation is more likely to happen in RIS 
systems.

%RIS 
%are more likely to appear the near-field effect,  
%	Nevertheless, we can conclude that the near-field range is proportional to 
%the square of the array aperture, and inversely proportional to the wavelength.

With the dramatically increased number of antennas and carrier frequency, the near-field range of ELAA considerably expands. {\color{black} For instance, we have recently fabricated a 0.36-meter-aperture ELAA at 28 GHz. If it is employed in SIMO/MISO scenarios, its near-field range is about 25 meters.} 
When both transmitter and receiver are equipped with this array, the near-field 
range becomes 100 meters. {\color{black} Moreover, if this ELAA works as a RIS 
with a BS-RIS distance of 50 meters, the near-field 
propagation should be accepted once the RIS-UE distance is shorter than 50 
meters.}
In summary, near-field communications come to be an indispensable part of 
future 6G. 
% In the following sections, we will elaborately discuss the recent studies of 
% near-field communication techniques.
%Take a 3200-element array working at 2.4 GHz as an example 
%\cite{RFocus_MIT2020}, its near-field range reaches 600 meters, covering a 
%large part of a standard cell. 

	\section{Challenges of Near-Field Communications}  \label{sec:3}
	The near-field propagation causes several challenges to wireless communications, i.e., existing 5G transmission methods specific for far-field suffer from severe performance loss in near-field areas. Technologies recently developed for addressing these challenges are discussed in this section.

	\subsection{Near-Field Channel Estimation}
	 \label{sec:3-1}

	\begin{figure}
	\begin{center}  
	\centering
	\hspace*{-5mm}
	\includegraphics[width=3in]{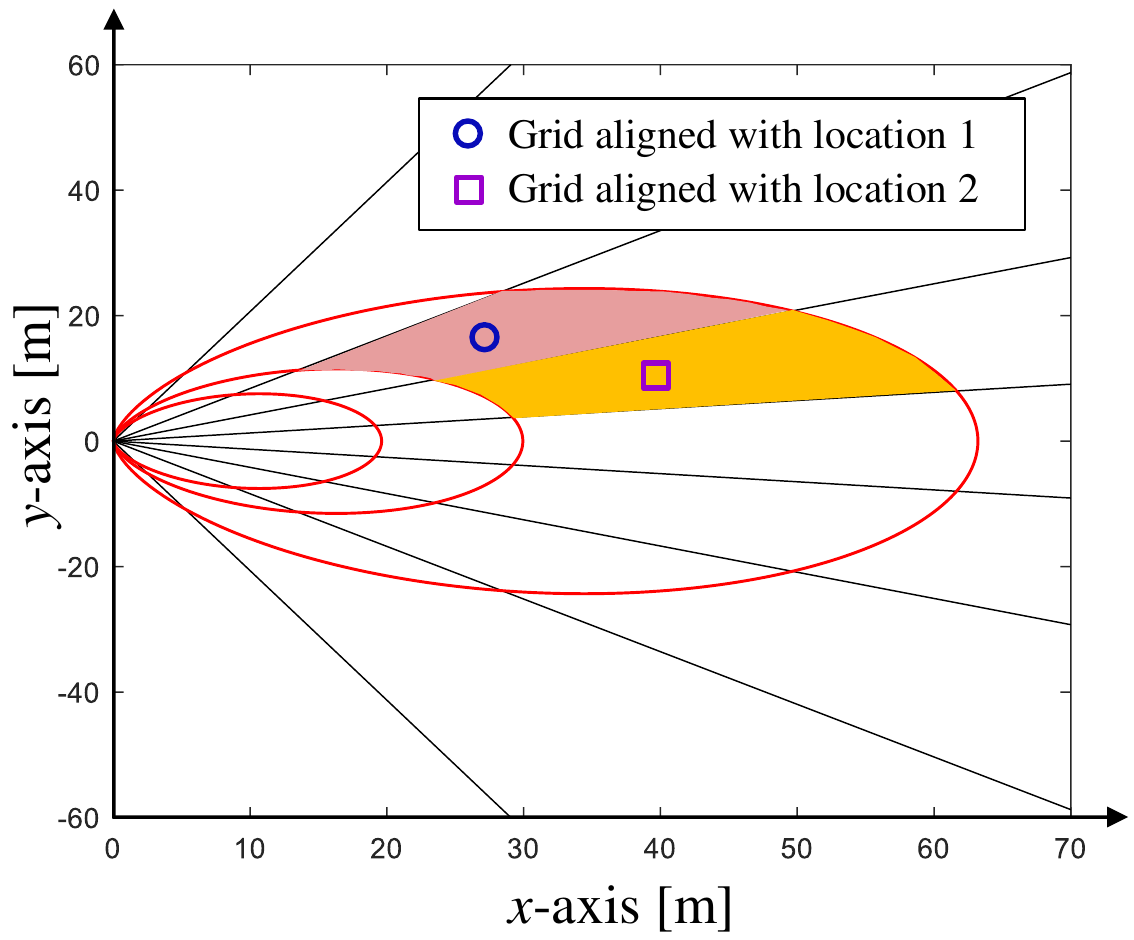}
	% \vspace*{-1em}
	\captionsetup{font = {small}}
	\caption{Near-field codebook with non-uniform grids.
	}
	\label{pic_CE}
	\vspace*{-1em}
	\end{center}
\end{figure}	 
				\begin{figure*}
	\centering
	% \hspace*{-5mm}
	\includegraphics[width=5.5in]{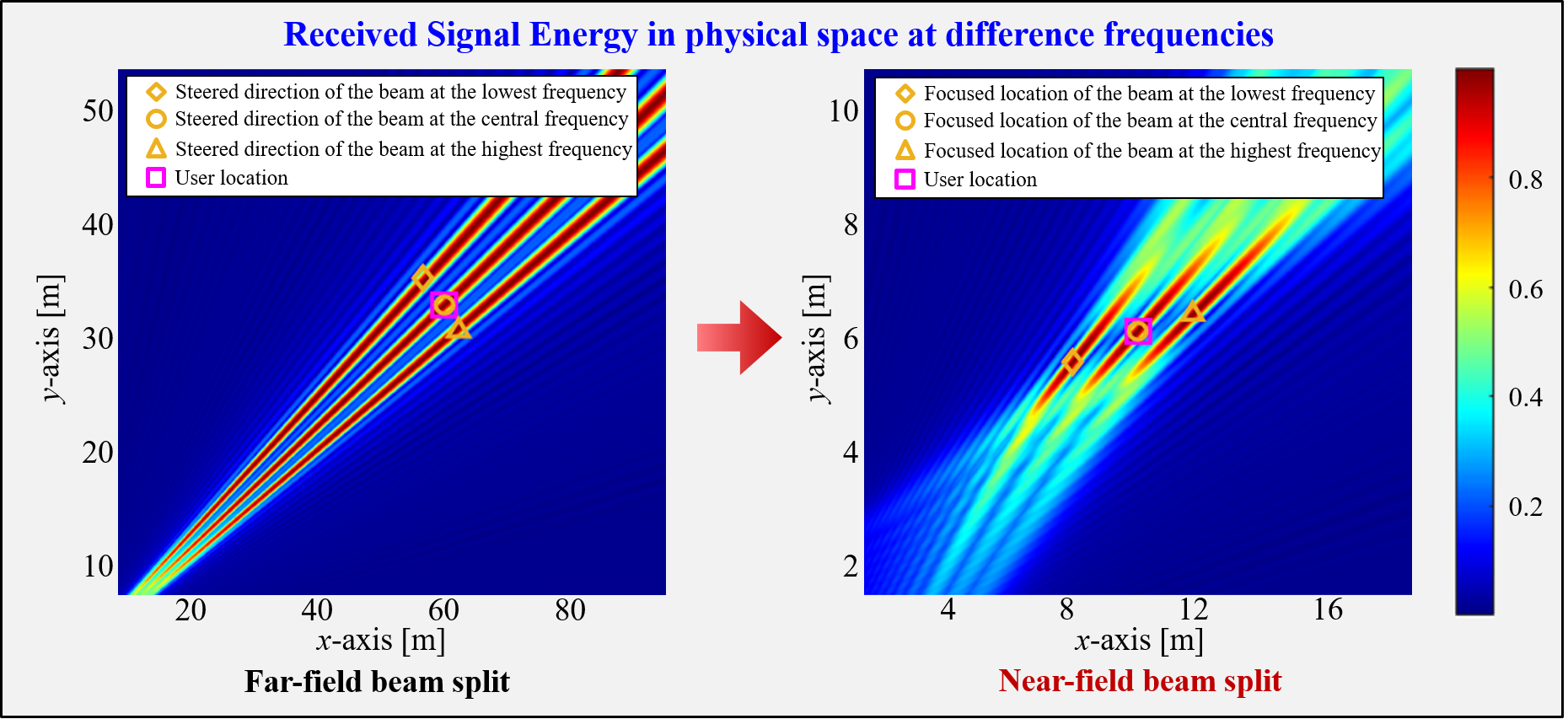}
	% \vspace*{-1em}
	\captionsetup{font = {small}}
	\caption{ \color{black}This figure illustrates the far-field beam split effect (left) and the near-field beam split effect (right). Far-field beam split makes beams at different frequencies transmit towards different directions, while near-field beam split makes beams at different frequencies be focused on various locations.  
	}
	\label{fig4}
	\vspace*{-1em}
\end{figure*}
	
		\emph{Challenge}: 
	Accurate channel estimation is required to attain the expected performance 
	gain of ELAA. As the number of channel paths is usually much smaller than 
	that of antennas, channel estimation methods with low pilot overhead 
	generally design suitable codebooks to transform the channel into a sparse 
	representation. For the far-field codebook, each
	codeword of the codebook corresponds to a planar wave
	associated with one incident angle. Ideally, each far-field path
	can be represented by \emph{only one} codeword. With this far-field
	codebook, the angle-domain representation of the channel can
	be obtained, and it is usually sparse due to the limited paths. Then, beam training and compressed sensing (CS) methods are applied to  estimate 
	far-field channels with low pilot overhead accurately. However, this far-field planar-wave codebook
	mismatches the actual near-field spherical-wave channel. This
	mismatch induces that a single near-field path should be jointly
	described by \emph{multiple} codewords of the far-field codebook.
	Accordingly, the near-field angle-domain channel is not sparse anymore, 
	which inevitably leads to the degradation of
	channel estimation accuracy. Therefore, near-field codebooks suitable for 
	near-field 
	channels 
	need to be carefully created.

		\emph{Recent progress}: 
		Some recent works have been endeavored to design near-field 
		codebooks 
		utilizing spherical wavefronts \cite{
		NearCE_Han2020, Near_Cui2022}.  
		In~\cite{NearCE_Han2020}, the entire two-dimensional physical 
		space is uniformly divided into multiple grids. Each grid is associated with a near-field array response vector, and all of these vectors construct the near-field codebook.
		With this codebook, the joint angle-distance information of each near-field path is extracted. Then, the near-field channel can be estimated by CS methods with low pilot overhead. 
However, with the decrease of BS-UE distance, the near-field propagation becomes dominant, and the distance information gradually becomes more crucial. 
Therefore, we can conceive the intuition that the grids should be sparse far 
away from the ELAA but dense near the ELAA. Without considering this intuition, the codebook in \cite{NearCE_Han2020} is hard to attain satisfactory channel estimation performance in the entire 
near-field region.  {\color{black} To this end, by minimizing the largest 
coherence among codewords in the near-field codebook, authors 
in~\cite{Near_Cui2022} mathematically prove this 
intuition}, i.e., the angle space could be uniformly divided, while the distance space should be non-uniformly divided. As shown in Fig.~\ref{pic_CE}, the shorter the distance,  the denser the grid. With the help of this non-uniform codebook, a 
polar-domain sparse channel representation and corresponding CS-based 
algorithms are proposed in~\cite{Near_Cui2022} to accomplish accurate channel estimation in both near- and far-field areas.

	\subsection{Near-Field Beam Split} \label{sec:3-2}

\emph{Challenge}: 
 In THz wideband systems, ELAA might encounter a beam split phenomenon, also 
 known as beam squint and spatial-wideband effect. { \color{black} Existing THz 
 beamforming architecture often employs analog phase-shifters (PSs) 
 \cite{Graphene_Singh2020}, which usually tune the same phase shift for signals 
 at different frequencies. Nonetheless, the actual phase of the  EM wave is the 
 product of the signal propagation delay and the \emph{frequency-dependent} 
 wavenumber. 
 As a result, the signal propagation delay can be compensated through a phase shift adequately only for a narrow band signal.
 % As a result, the phase of radiated signals can only be well 
 % compensated at one frequency by PS. 
 Phase errors are introduced for the other 
 frequencies, thus causing the beam split effect. In fact, the impact of beam 
 split on far-field and near-field propagations also differs.
 
  	In far-field, beam split leads to the fact that beams at different 
 frequencies are transmitting towards different angles, as shown in the left 
 figure of Fig. \ref{fig4}. For near-field beam split, however, beams are 
 focused at both different angles and various distances due to the split of 
 spherical waves,  as shown in the right 
 figure of Fig. \ref{fig4}. Both far-field and near-field beam splits severely 
 reduce the received signal energy of frequency components misaligned with the 
 user location.
 Over the years, extensive works have been proposed to mitigate far-field beam 
 split by tuning \emph{frequency-dependent} phase shifts with planar wavefronts 
 through 
 true-time-delay-based (TTD-based) beamforming instead of PS-based 
 beamforming. Unfortunately, owing to the discrepancy between planar and 
 spherical 
 waves, these schemes addressing the far-field beam split no longer work 
 well in the near-field, posing challenges to THz ELAA 
 communications.}
	
% 	In far-field, beam split leads to the fact that beams at different 
% 	frequencies are transmitting towards different angles, and the received 
% 	signal energy is severely
% 	reduced, as shown in the left figure of Fig. \ref{fig4}. Over the years, 
% 	extensive works have been proposed to mitigate far-field beam split by 
% 	tuning 
% 	\emph{frequency-dependent} phase shifts with planar wavefront through 
% 	true-time-delay-based (TTD-based) beamforming instead of PS-based 
% 	beamforming. For near-field beam split, however, beams are focused at both 
% 	different angles and various distances,  as shown in the right figure of 
% 	Fig. \ref{fig4}. That is because of the split of spherical waves. Owing to 
% 	the difference between planar and spherical waves, existing schemes 
% 	addressing the far-field beam split effect cannot work well anymore in the 
% 	near-field, which poses challenges to THz ELAA communications.}

\emph{Recent progress}: Recently, a few efforts have tried to overcome 
the near-field beam split effect. {\color{black}
In \cite{NearBF_Health2021}, a variant of chirp sequence is utilized to design 
the phase shifts, for flattening the beamfocusing gain across frequencies 
with the sacrifice of the maximum beamfocusing gain. 
This method can slightly alleviate the near-field beam split effect, but its  spectral efficiency degrades as well when the bandwidth is very large, as the beams are still generated by PSs. }
To this end,  a phase-delay focusing (PDF) method is proposed in \cite{NearBF_Cui2021} exploiting TTD-based beamforming. 
 To further illustrate, the BS ELAA is first partitioned into multiple 
 sub-arrays. The UE is assumed to be located in the far-field area of each 
 small sub-array but within the near-field range of the ELAA. Then, one TTD 
 line is inserted between each sub-array and the radio-frequency (RF) chain to 
 realize frequency-dependent phase shifts. Finally, the frequency-dependent 
 phase variations across different sub-arrays induced by spherical wavefronts 
 are compensated by the inserted TTD line. As a result, beams over the working 
 band are focused at the target UE location \cite{NearBF_Cui2021}. 

{\color{black} In conclusion, the first solution \cite{NearBF_Health2021} follows 
the  PS-based beamforming, which is easy to implement but the achievable 
performance is unsatisfactory. The second scheme \cite{NearBF_Cui2021} can 
nearly eliminate the near-field beam split effect but requires the 
implementation of TTD lines. 
In fact, although deploying TTD lines by optical fibers has been demonstrated 
in the optical domain, this kind of deployment is non-trivial to be extended to 
THz ELAA communications. Fortunately, recent advances in graphene-based 
plasmonic waveguides provide low-cost solutions for implementing TTD lines at 
high frequencies \cite{Graphene_Singh2020}.
}

{
\section{Potentials for Near-Field Communications} \label{sec:4}

Unlike the aforementioned works for dealing with the performance degradation in the near-field, some recent studies have surprisingly 
revealed that 6G networks can also benefit from near-field propagation. 
In this section, we will discuss those studies exploiting the potentials of near-field propagation to improve communication performance.

\subsection{Capacity Enhancement}\label{sec:4-1}
\emph{Potential}: 
{\color{black} The spatial multiplexing gain of MIMO communications considerably 
increases with the transition from far-field regions to near-field regions. In 
far-field MIMO communications, the line-of-sight (LoS) channel can be 
represented by a rank-one 
matrix, where the spatial degrees-of-freedom (DoFs) are very limited. By 
contrast, the near-field LoS channel can be rank-sufficient  derived from the 
geometric relationship under the spherical propagation model. The increased 
rank 
indicates dramatically improved spatial DoFs in the near-field region. }
Precisely, based on the expansion of prolate spheroidal wave functions, it is 
proved in \cite{DoF_Miller2019} that 
near-field spatial DoFs are proportional to the product of the BS and UE array 
apertures and inversely proportional to the BS-UE distance. 
This conclusion is further improved in \cite{DoF_Nicolo2021} by meticulously 
designing the beamfocusing vectors of the BS and UE arrays.
As shown in Fig. \ref{fig5}, the DoFs increase from $1$ to 
$20$ when the BS-UE distance decreases from $350$ meters to $10$ 
meters. {\color{black} Thanks to the increased DoFs, the near-field LoS path enables simultaneous transmission of multiple data streams by MIMO precoding, as opposed to the rank-one far-field LoS channel supporting only one data stream. 
The increased spatial DoFs can be exploited as an additional spatial multiplexing gain, which offers a new possibility for a significant capacity enhancement.}

\begin{figure}[!t]
	\centering
	% \hspace*{-5mm}
	\includegraphics[width=3.5in]{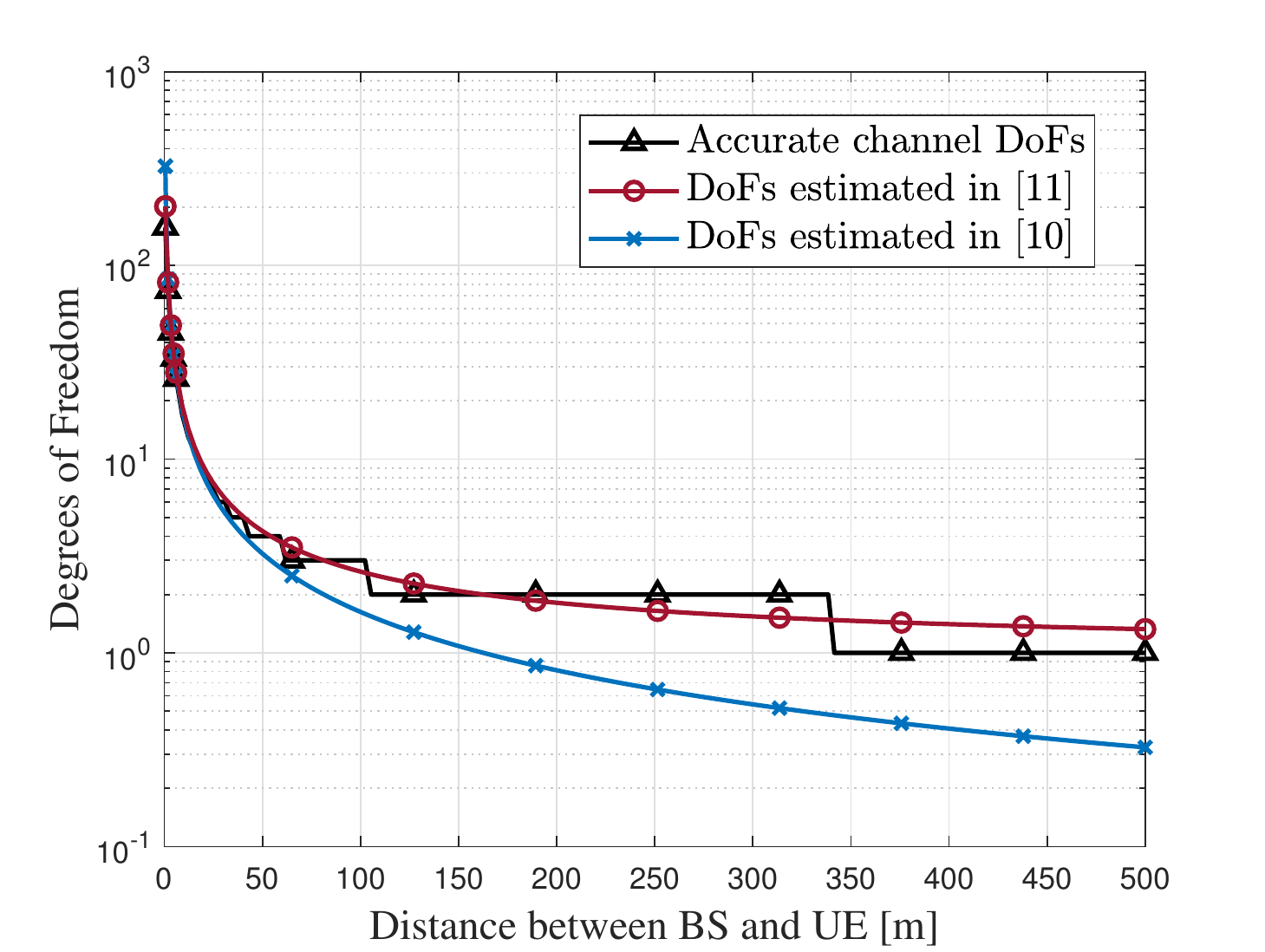}
	\captionsetup{font = {small}}
	\caption{The spatial DoF increases in the near-field region.}
	\label{fig5}
	\vspace*{-3mm}
\end{figure}
\par 
\emph{Recent progress}: 
{
% Recently, some novel precoding architectures have proposed to leverage these extra near-field DoFs for MIMO capacity enhancement \cite{NearHP_Wu2021, NearHP_Yan2022}. 
% Firstly, hybrid precoding architecture with reduced RF chains is widely considered in 5G systems owing to its low power consumption. Unfortunately, since the number of RF chains of hybrid precoding (e.g., 2 or 4 RF chains) is much smaller than the increased near-field DoFs (e.g., DoFs are 20 at 10 meters), classical hybrid precoding cannot fully benefit from the near-field propagation. For exploiting the extra near-field DoFs, a distance-aware precoding (DAP) architecture is developed in \cite{NearHP_Wu2021}. Unlike the hybrid precoding with a fixed and limited number of RF chains, the DAP architecture could flexibly adjust the number of RF chains to match the distance-aware DoFs by configuring each RF chain as active or inactive. 
% The DAP architecture could significantly increase the spectral efficiency with similar energy efficiency compared with hybrid precoding. {\color{black} Another effort to harvest the additional spatial multiplexing gain provided by the spherical wave propagation is the widely-spaced multi-subarray (WSMS) hybrid precoding \cite{NearHP_Yan2022}. In this architecture, the sub-arrays are widely spaced to enlarge the array aperture, therefore creating the artificial expansion of the near-field region. With elaborate optimization of the number and spacing of the sub-arrays, the WSMS architecture could achieve spectral efficiency superior to hybrid precoding. 
Recently, some novel precoding architectures have been proposed to leverage these extra near-field DoFs for MIMO capacity enhancement \cite{NearHP_Wu2021, NearHP_Yan2022}. 
Firstly, distance-aware precoding (DAP) is developed in \cite{NearHP_Wu2021}. Unlike classical hybrid precoding with a fixed and limited number of RF chains (e.g., 2 or 4 fixed RF chains), the DAP architecture could flexibly adjust the number of RF chains to match the distance-related DoFs, which is achieved by deploying a selection network to configure each RF chain as active or inactive. For instance, in the far-field region, only one RF chain is activated for data transmission. When communication distance is decreasing to 10-20 meters, around 20 activated RF chains are enough to 
adapt to the DoFs, as shown in Fig. \ref{fig5}. By doing so, the number of 
transmitted data streams dynamically matches the DoFs. Simulations demonstrate 
the DAP could significantly increase the spectral efficiency while its energy 
efficiency is comparable with hybrid precoding. {\color{black} To avoid the 
utilization of extra RF chains, another effort to harvest the potential spatial 
multiplexing gain in near-field areas is the widely-spaced multi-subarray 
(WSMS) precoding \cite{NearHP_Yan2022}.
In this architecture, the sub-arrays are widely spaced to enlarge the array aperture, artificially creating the expansion of the near-field region. Compared with classical hybrid precoding, the number of sub-arrays and the sub-array spacing should be additionally designed in the WSMS architecture.
To this end, \cite{NearHP_Yan2022} first assumes planar-wave propagation within 
each sub-array and spherical-wave propagation across different sub-arrays 
similar to \cite{NearBF_Cui2021}. Then, \cite{NearHP_Yan2022} jointly 
optimizes 
the number of sub-arrays, their spacing, and the precoding matrix for 
maximizing the achievable rate.  Simulations demonstrate that WSMS could 
achieve nearly 200\% higher spectral efficiency than classical hybrid precoding.

% The joint optimization of the array configuration and hybrid precoding can be decoupled into two sub-problems. To design the number of the sub-arrays and their spacing, a dominant-LoS relaxation (DLR) algorithm is proposed, which utilizes the dominant LoS channel to represent the complicated multi-path channel to simplify the optimization process. Then a close-form solution of hybrid precoding can be obtained by assuming planar propagation within each sub-array and spherical propagation within different sub-arrays. By elaborate optimization, WSMS could achieve nearly 200\% higher spectral efficiency compared with classical hybrid precodings.

}

}

\subsection{Accessibility Improvement}\label{sec:4-2}
\begin{figure*}
	\centering
	% \hspace*{-5mm}
	\includegraphics[width=6.5in]{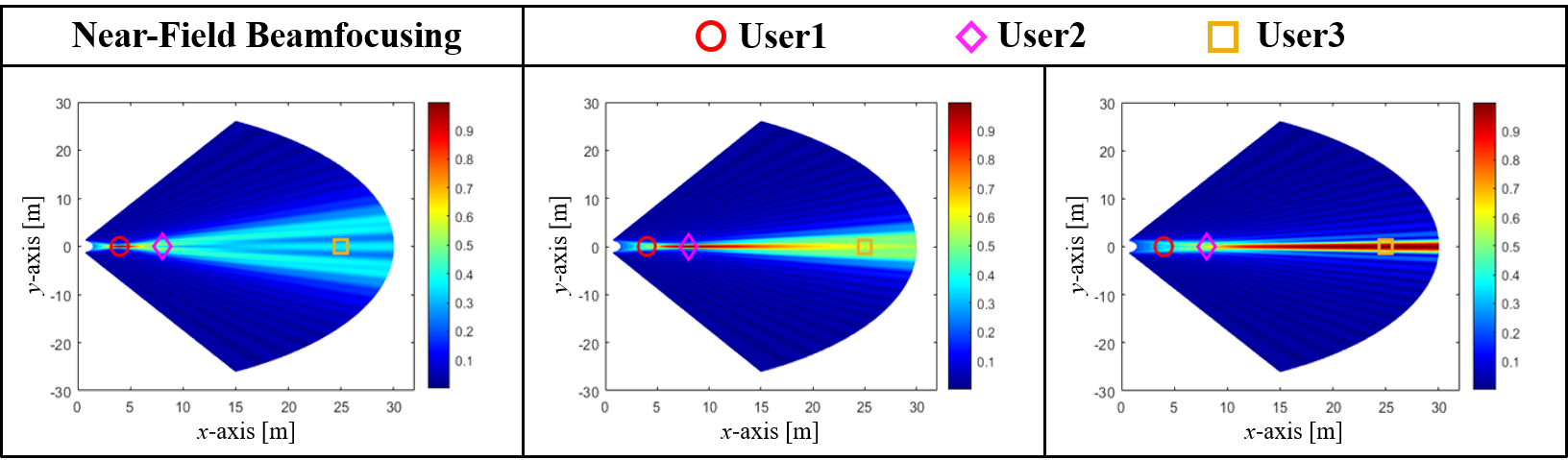}
	\captionsetup{font = {small}}
	\caption{Near-field beamfocusing is able to serve multiple users 
	in the same angle.}
	\label{fig6}
	\vspace*{-3mm}
\end{figure*}

\emph{Potential}: Near-field propagation is also able to improve accessibility 
in multi-user (MU) communications. 
To increase the spectral efficiency in MU-MIMO communications, space division 
multiple access (SDMA) is widely considered to distinguish users 
through orthogonal or near-orthogonal spatial beams. Thus, multiple users can 
share the same time and frequency resources. 
For far-field SDMA, utilizing beamsteering to generate beams with planar 
wavefronts can distinguish users at different angles. 
A downside is that users located at similar angles will severely interfere with 
each other, and thus can not simultaneously access the network through 
far-field SDMA. 
Fortunately, near-field beamfocusing enjoys the capability of energy focusing 
on the joint angle-distance domain. Hence, near-field SDMA could generate 
beams with spherical wavefronts to simultaneously serve users located at
similar angles but different distances, as shown in Fig. \ref{fig6}. The 
distance information of 
spherical wavefronts supplies a new utilizable dimension for multi-user access, 
which is not achievable for conventional far-field SDMA.

\emph{Recent progress}: Taking advantage of the capability of beamfocusing, the 
authors 
in \cite{NearFocus_Zhang2022} have studied the near-field multi-user 
transmission considering fully-digital precoding, hybrid precoding, and 
transmissive reconfigurable metasurface (RMS). 
 By optimizing the sum rate in multi-user systems through alternating 
 optimization, all considered precoding architectures can naturally generate 
 beams with spherical 
 wavefronts to distinguish users located at 
 similar angles but different distances. 
 The simulation results demonstrate that near-field propagation has the 
 potential of 
 enhancing multi-user accessibility.

	}
	\section{Future Research Directions} \label{sec:5}
	In this section, several future research directions for 
	near-field 	communications will be pointed out.
	\subsection{Near-Field Communication Theory} \label{sec:5-1}
% 	\subsubsection{Near-Field Capacity Analysis}
% {It has been illustrated in \cite{DoF_Miller2019, DoF_Nicolo2021} that, the 
% number of supportable data streams (spatial DoFs) could be accurately 
% estimated, and thus a theoretical analysis of the near-field channel capacity 
%  can be obtained. However, the derivation is 
% restricted to parallel arrays, where the array depth along the transmission 
% direction is neglected. When it comes to more practical non-parallel arrays, 
% the estimated DoFs may be not accurate any more. 
% Moreover, these analyses \cite{DoF_Miller2019, DoF_Nicolo2021} mainly consider 
% the uniform linear array (ULA) and uniform planar array (UPA), while other 
% array structures, such as sparse array and RIS, have not 
% been investigated. 
% Therefore, a DoF estimation method and the 
% corresponding near-field capacity analysis for generalized arrays  are 
% required. }
	
{\color{black}	\subsubsection{Improvement of Rayleigh Distance}
	As a widely adopted quantification of near-field range, Rayleigh distance is attained in terms of phase discrepancy.
	For communication metrics directly affected by phase discrepancy, such as 
	channel estimation accuracy,  Rayleigh distance can accurately capture the 
	degradation of these metrics when applying far-field transmission schemes 
	in the near-field region.
	On the contrary, some metrics are directly influenced by other factors instead of phase discrepancy, e.g., capacity is determined by beamforming gain and channel rank. Accordingly, classical Rayleigh distance probably cannot capture the performance loss of these metrics well.
	To this end, several recent works have endeavored to improve classical Rayleigh distance in terms of some vital communication metrics. 
	For instance, an effective Rayleigh distance (ERD) is derived in \cite{NearBF_Cui2021} for the accurate description of beamforming gain loss and capacity loss.
	Nevertheless, ERD is only valid for MISO communications, while more 
	discussion should be made to improve Rayleigh distance in more practical 
	scenarios under more general metrics, e.g., channel rank and energy 
	efficiency in MIMO and RIS systems.
}
%	2. Channel model

%	1. Distance-aware precoding + sub-array design + Lens, RIS [Mingyao Cui]

	\subsection{Near-Field Transmission Technologies} \label{sec:5-3}
	\subsubsection{AI-Aided Near-Field Communications} 
Different from far-field communications, the transmission algorithms for 
near-field communications are more complex. To be specific, since extra grids 
on the distance domain are required, as mentioned in Section \ref{sec:3-1} 
\cite{Near_Cui2022}, the size of near-field codebooks is usually much larger 
than that of far-field codebooks, leading to high-complexity channel estimation 
and beamforming. Moreover, the \emph{non-linear} phase characteristics of 
spherical waves make the design of near-field beam training and precoding 
algorithms more complicated than that in far-field areas. 
AI-based transmission methods are promising to address these problems since 
they can mine the features of  near-field environments through 
\emph{non-linear} neural networks. {\color{black} Currently, there are plenty of 
works elaborating on AI-based far-field transmissions (references are not 
provided here since the number of references is limited in this magazine), 
while AI-based 
near-field transmissions have not been well studied.
}

	\subsubsection{RIS-Aided Near-Field Communications} 
{
	Compared with MIMO communications, the near-field propagation becomes even more dominant and complex in RIS-aided systems. 
	In MIMO communications, based on the spherical propagation model, the 
	EM waves form \emph{spherical} equiphase surfaces 
	at the receiver. 
	On the contrary, in RIS-aided systems, the phase of received EM waves is accumulated by the propagation delays through the BE-RIS and RIS-UE links. Based on the geometry relationship, the equiphase surfaces become \emph{ellipses} in the near-field range instead of spherical.
	{\color{black} Accordingly, the research on beamfocusing \cite{RISFocus_Dovels2021}, channel estimation, and multiple access techniques taking into account this ellipses-equiphase property are required for RIS-aided near-field communications.}
}

	\subsubsection{Hybrid Far- and Near-Field Communications} 
	{\color{black}
	In practical systems, communication environments usually exist with both far-field and near-field signal components. 
	First, in multi-user systems with multi-path channels, 
	some users and scatterers may be far away from the BS while others are located in the near-field region of the BS, which constitutes a hybrid far- and near-field (hybrid-field) communication scenario.
	Additionally, it is worth mentioning that the Rayleigh distance is proportional to frequency. Thus, in an ultra-wideband or frequency-hopping system with a very large frequency span, its near-field range varies dramatically across the bandwidth. Chances are that the signal components at low frequencies may operate in far-field regions while those at high frequencies with larger Rayleigh distances are propagating in the near-field areas, which also contributes to hybrid-field communications. 
	Consequently, the above factors make hybrid-field communications practical and crucial in future 6G networks. Thus, hybrid-field transmission techniques handling both far-field and near-field signal components deserve in-depth study. 
	
% 	need to be elaborately modeled and analyzed. Although near-field techniques may also work well for hybrid-field communications, they usually require high computational complexity for the far-field components in hybrid-field channels. Thus, the specialized  hybrid-field transmission techniques with low complexity are worth studying for 6G ELAA.
	}
	
		\subsubsection{Spatial Non-Stationarity Effect on Near-Field Communications} 
	{\color{black}
Except for near-field propagation, the spatial non-stationarity effect is 
another fundamental characteristic of ELAA compared to 5G massive MIMO, where 
different scatterers and users are visible to different portions of the ELAA. 
This effect leads to the fact that only a part of the ELAA can receive the 
spherical EM waves radiated by a scatterer or a user. The angular power 
spectral and average received power rapidly vary over the ELAA.
Recently, there have been intensive works dealing with the non-stationarity effect and near-field propagation simultaneously \cite{NearCE_Han2020}. However, the impart of non-stationarity on other emerging near-field communications has not been well studied, such as RIS-aided systems and hybrid-field communications.}

	\subsection{Hardware Development} \label{sec:5-2}
	To verify the effectiveness of near-field transmission technologies, hardware developments and over-the-air experiments are of great significance. 
	For example, for alleviating the near-field beam split effect, TTD lines 
	need to be meticulously designed in the THz domain. The hardware 
	developments of WSMS and DAP architectures are worth being carried out to 
	exploit the near-field spatial DoFs. 
    Besides, implementing these techniques still has to overcome several hardware impairment issues, including In-phase/Quadrature imbalance, low-efficiency power amplifier at high frequency, etc. All these challenges should be carefully addressed to enable the implementation of 6G near-field communications.

% The challenges and potentials of near-field communications require some changes 
% of  the array architecture. 
% For example, as we have shown in Subsections \ref{sec:3-2} that, 
% time-delay circuits are introduced to address the near-field beam split effect. 
% For taking advantage of the distance-aware near-field DoFs, DAP architecture is 
% utilized in Subsection \ref{sec:4-1}. To further exploit the near-field effect, 
% more array architecture 
% designs may be required in the near future. For instance, for realizing 
% near-field transmissions, the focal point of a lens array should be delicately 
% designed.
% 	\subsection{Hardware Verification} \label{sec:5-4}
% 	{
% To verify the effectiveness of the aforementioned near-field transmission 
% technologies, the  over-the-air hardware experiments are very 
% important for future research. In ELAA communication systems, the 
% near-field channel parameters could be measured to verify the near-field 
% communication theory. 
% The experiments for evaluating the performances of 
% near-field transmission techniques, e.g., the DoF analysis results, are worth 
% carrying out to prove the necessity of near-field ELAA research. }

	\section{Conclusions} \label{sec:6}
{\color{black}With the evolution from massive MIMO to ELAA, near-field 
propagation with spherical wavefront becomes indispensable in 6G networks, 
where conventional far-field propagation with planar wavefront is not valid 
anymore. In this article, we revealed that near-field propagation is a 
double-edged sword, i.e., it brings both challenges and potentials to 6G 
communications. We first introduced the non-linear phase property of spherical 
waves and explained the derivation of near-field range in terms of phase 
discrepancy. Then, we discussed the technical challenges of channel estimation 
and beam split caused by near-field propagation and presented the recent 
solutions. In addition, some appealing works that exploit the capability of 
spherical waves to improve capacity and accessibility were
investigated. 
Several future research directions for near-field communications, such as 
improvement of Rayleigh distance and hybrid-field transmissions, were also 
highlighted, which are expected to inspire more innovations on 6G near-field 
communications.}

	\section*{Acknowledgement}
This work was supported in part by the National Key Research and Development Program of China (Grant No.2020YFB1807201), in part by the National Natural Science Foundation of China (Grant No.62031019). 

	% \balance
% % 	\bibliographystyle{IEEEtran}
\bibliographystyle{IEEEtran}
\bibliography{refs, IEEEabrv}

	\section*{Biographies}
		{\textbf{Mingyao Cui}} is a
		M.S. researcher in BNRist from Tsinghua University, 
		Beijing, China.
		\par \quad \par
		{\textbf{Zidong Wu}} is a Ph.D. researcher in BNRist from Tsinghua 
		University, Beijing, China.
		\par \quad \par
		{\textbf{Yu Lu}} is a Ph.D. researcher in BNRist from Tsinghua 
		University, Beijing, China.
	\par \quad \par
		{\textbf{Xiuhong Wei}} is a M.S. researcher in BNRist from Tsinghua   
		University, Beijing, China.
		\par \quad \par
		{\textbf{Linglong Dai}} is an associate professor at Tsinghua University. His current research interests include RIS, massive MIMO, mmWave and THz communications, and machine learning for wireless communications. He has received six conference best paper awards and four journal best paper awards.
		
		% {\textbf{Linglong Dai (Fellow, IEEE)} received the Ph.D. degree (with the highest honor) from Tsinghua University, Beijing, China, in 2011. From 2011 to 2013, he was a Postdoctoral Research Fellow with the Department of Electronic Engineering, Tsinghua University, where he was an Assistant Professor from 2013 to 2016 and has been an Associate Professor since 2016. His current research interests include RIS, massive MIMO, mmWave and THz communications, and machine learning for wireless communications. He has received six conference best paper awards and four journal best paper awards.
		% }

\end{document}